\documentclass[conference]{IEEEtran}

\IEEEoverridecommandlockouts
\usepackage{tikz}
\usepackage{cite}
\usepackage{amsmath,amssymb,amsfonts}
\usepackage{algorithmic}
\usepackage{graphicx}
\usepackage{textcomp}
\usepackage{xcolor,float}
\usepackage{url,booktabs}
\usepackage{comment}

\def\BibTeX{{\rm B\kern-.05em{\sc i\kern-.025em b}\kern-.08em
    T\kern-.1667em\lower.7ex\hbox{E}\kern-.125emX}}
    
\newcommand\copyrighttext{%
  \footnotesize \textcopyright 2021 IEEE. Personal use of this material is permitted.
  Permission from IEEE must be obtained for all other uses, in any current or future
  media, including reprinting/republishing this material for advertising or promotional
  purposes, creating new collective works, for resale or redistribution to servers or
  lists, or reuse of any copyrighted component of this work in other works.}

\begin{document}

\title{A Computer Science-Oriented Approach to Introduce Quantum Computing to a New Audience}

\author{
\IEEEauthorblockN{Özlem Salehi}
\IEEEauthorblockA{\textit{Department of Computer Science} \\
\textit{Özyeğin University}\\
İstanbul, Turkey \\
ozlemsalehi@gmail.com}
\and 
\IEEEauthorblockN{Zeki Seskir}
\IEEEauthorblockA{\textit{Department of Physics} \\
\textit{METU}\\
Ankara, Turkey \\
zeki.seskir@gmail.com}
\and
\IEEEauthorblockN{İlknur Tepe}
\IEEEauthorblockA{\textit{} \\
\textit{QWorld} \\
ilknrtepe@gmail.com
}
}

\newcommand\copyrightnotice{%
\begin{tikzpicture}[remember picture,overlay]
\node[anchor=south,yshift=10pt] at (current page.south) {\fbox{\parbox{\dimexpr\textwidth-\fboxsep-\fboxrule\relax}{\copyrighttext}}};
\end{tikzpicture}%
}

\maketitle
\copyrightnotice 
\begin{abstract}
Contribution: In this study, an alternative educational approach for introducing quantum computing to a wider audience is highlighted. The proposed methodology considers quantum computing as a generalized probability theory rather than a field emanating from physics and utilizes quantum programming as an educational tool to reinforce the learning process.

Background: Quantum computing is a topic mainly rooted in physics, and it has been gaining rapid popularity in recent years. A need for extending the educational reach to groups outside of physics has also been becoming a necessity.

Intended outcomes: This study aims to inform academics and organizations interested in introducing quantum computing to a diverse group of participants on an educational approach. It is intended that the proposed methodology would facilitate people from diverse backgrounds to enter the field

Application design: The introductory quantum physics content is bypassed and the quantum computing concepts are introduced through linear algebra instead. Quantum programming tasks are prepared in line with the content. Pre/post-test design method and Likert scale satisfaction surveys are utilized to measure knowledge acquisition and to evaluate the perception of the learning process by the participants.

Findings: Conducted pre/post-test design survey shows that there is a statistically significant increase in the basic knowledge levels of the participants on quantum computing concepts. Furthermore, no significant difference in the gain scores is observed between the participants from different STEM-related educational backgrounds. The majority of the participants were satisfied and provided positive feedback. 
\end{abstract}


\section{Introduction}

The idea of a quantum computer was first popularized by Richard Feynman near 40 years ago \cite{feynman82}. The concept of a second quantum revolution expanded this idea to an array of quantum technologies \cite{dowling} that has been utilized for almost two decades now. The developments in the hardware realm are also supported by national initiatives in the order of billion dollars by countries like UK \cite{q_uk}, China \cite{q_ch}, Russia \cite{q_rus}, and the US \cite{q_us1,q_us2}. Also on a regional level, the Quantum Flagship program of the European Union is the largest international initiative of this scale \cite{q_eu}. Most of these initiatives have a strong focus on not only research and development but also on the commercialization of these technologies. This focus on commercialization and maturation of quantum technologies allowed hundreds of firms to start exploring the commercial opportunities presented by these developing set of technologies. As of 2020, there are more than 100 firms focused on quantum computing\footnote{https://quantumcomputingreport.com/privatestartup/, Accessed: May 4, 2021}. It has been noted in the literature that expanding the general level of understanding on quantum technologies, especially quantum computing, is gaining importance for both societal impact \cite{vermaas2017}, and commercialization purposes in terms of developing the necessary quantum workforce \cite{nap2019}.

Despite the necessity of reaching a wider audience, quantum computing courses are usually offered in physics and computer science departments at the graduate level. Most of the courses begin with teaching the concepts from quantum physics and focus on the theoretical aspects, which make quantum computing a relatively harder subject to grasp. 

In this study, the aim is to share a methodology with academics and organizations interested in introducing quantum computing to a wider audience. The proposed approach introduces quantum computation as a generalization of probabilistic computing through task-based quantum programming. Viewing quantum computation as a generalization of probabilistic computing is an approach adopted by computer scientists \cite{YM08, AB09} and popularized by Scott Aaronson in his book ``Quantum Computing since Democritus'' \cite{aaronson2013}. Leaving the underlying physics behind, this approach allows teaching quantum computing phenomena through abstract and linear algebraic concepts. Note that the prerequisites for this approach are basic knowledge in linear algebra and introductory programming experience, which are already met by most of the undergraduate students from STEM-related departments. 

Following this methodology, an introductory material\footnote{The material, called Bronze, can be accessed here: https://qworld.net/workshop-bronze/. Accessed: May 4, 2021} is developed by the QWorld initiative, which is intended for everyone with a background in linear algebra and basic programming experience in Python. Quantum programming is used as the tool to introduce quantum computing, and the task-based programming approach helps participants to learn through experience. Altogether, this methodology allows an alternative approach through a skills-based learning process for those not interested in the physics of quantum computing, as well as helping participants to gain basic quantum programming skills. 

To this end, the main objectives of this study are to demonstrate the viability of this approach by assessing the knowledge acquisition and satisfaction levels of the participants from a diverse background through pre/post-test surveys conducted during the workshops and to introduce this topic into the research literature to allow a further investigation to build upon the preliminary empirical findings presented. 

\section{Background}\label{sec: Back}

This section introduces the previous and current studies in the literature on quantum computing education, why it is becoming a viable topic for scholarly inquiry, several small scale empirical works that can be considered as a precursor to the work presented in this paper, and other educational materials on quantum programming that can be utilized for further studies.

A recent study in which 21 companies from the quantum industry are interviewed reveals that coding is the most valued skill according to these employers \cite{FZL20}. The study further suggests that an introductory level quantum course focusing either on the hardware or software track to increase quantum awareness would be useful for students from different science majors to enter into the field. A similar point is made by \cite{Pe20}, which emphasizes the need for training of engineers and programmers as specialists in quantum programming and software to prepare for the quantum workforce. Anticipating these rapid developments, in 2017, a diverse group of researchers from academia and industry published a document titled ``Quantum Software Manifesto'' \cite{qusoft_manf}. This document emphasizes "educating more quantum programmers," among other crucial points.  Another study on quantum engineering conducted with 26 experts in the field emphasizes the need for increasing ``quantum awareness'' for the quantum workforce of the future \cite{gerke2020}, whereas arguing that this is most necessary for non-physicists. Similarly, concepts like quantum literacy \cite{Laur}, are introduced as reaching wider audiences becomes important.

The literature demonstrates that quantum computation is still more closely related to quantum information theory concepts than experimental fields on a conceptual level of mapping for quantum technologies-related academic topics, but the academic literature is overwhelmingly rooted in physics \cite{seskir2019}. The topic of how to introduce quantum computing to a wider audience (especially computer scientists) has been around for almost two decades \cite{Me03}, and new methods that aim to make quantum computing more available for high-school and early undergraduate students are being developed \cite{barnes2020,Quest}. 

There are several studies reporting student and teacher experience in learning and teaching quantum physics \cite{Sin01, Ire00, KPB17, Me03, MW02}. In \cite{Aka10}, the main difficulties in teaching quantum physics as identified by the professors are listed as lack of knowledge of mathematical concepts necessary for quantum physics and insufficient background in classical mechanics. Apart from mathematical difficulties, another problem stems from abstract physical concepts as quantum physics is less intuitive when compared to other physics courses. The computer science-oriented approach which views quantum computing as a generalization of probabilistic computing has also been used in many lectures and informal educational materials. Yet, no studies discussing this methodology on an empirical level with a wide range of participants with different backgrounds were found in the literature. 

As quantum computing is becoming a popular subject, the number of courses and educative events is increasing. In 2019, Microsoft prepared an introductory quantum computing course that focused on the implementation of quantum algorithms by Microsoft's quantum programming language Q\#. Following the first course offered at the University of Washington, Microsoft paired with 10 more institutions around the world \cite{Microsoft}. Likewise, IBM has prepared the Qiskit Textbook which is a collection of Jupyter Notebooks to introduce quantum computing through Qiskit, a collection of open-source Python libraries mainly developed by IBM \cite{Qiskit}. The textbook can be used for self-study or by educators to integrate quantum programming into existing quantum computing curricula.

As courses and events related to quantum computing grow in numbers, the studies discussing experiences in teaching quantum computing are appearing as well. A report about the Microsoft course offered at the University of Washington is prepared by \cite{MS20}, based on the observations of the authors during the course, the performance of the students, and the feedback forms completed by 14 students out of 35 who completed the course. Overall, the study puts forward that a software-oriented approach is a viable way to introduce students to quantum computing. \cite{HIPST20} discuss the survey results of 45 participants who took the course ``Quantum Computing as a High School Module'' prepared \cite{PSHIT19} for high school students between the age 15 and 18. The students have been successful at acquiring the new concepts and the course was able to generate a baseline knowledge on quantum computing. \cite{larose2019} discusses experiences from a one-week quantum computing course relying on quantum programming and aimed at students with no physics background. The survey results show that the students got the base for learning more on their own and most of them were interested in learning more. \cite{carrascal2020} presents a methodology based on quantum programming for introducing undergraduate students with no physics background to quantum computing and shares the survey results which supports that of \cite{larose2019}.

\section{Methods}\label{sec: bro}

\subsection{Approach}

 The educational material utilized in this study is designed to teach quantum computing from a linear algebra and computer science perspective, by presenting quantum computing as a generalization of probabilistic and classical computing and abstracting the physical concepts through linear algebra. This is discussed elegantly in \cite{aaronson2013} through the following quote:

\begin{quote}
    “Quantum mechanics is what you would inevitably come up with if you started from probability theory, and then said, let’s try to generalize it so that the numbers we used to call “probabilities” can be negative numbers. As such, the theory could have been invented by mathematicians in the nineteenth century without any input from experiment. It wasn’t, but it could have been.”
\end{quote}

This approach has several distinguishing features against the traditional approach. In traditional quantum computing courses, the concept of superposition (the property of a quantum state to be in multiple basis states simultaneously) follows its historical roots, either through photon polarization or the spin of an electron. For the photon polarization case, the concept of light as an electromagnetic wave with propagation at a certain direction, and polarization in an orthogonal direction needs to be introduced. This requires a diversion into the theory of light and how waves propagate, which are not obvious for participants without a background covering at least one semester of modern physics. Similarly, introducing superposition using the spin of an electron is discussed in physics courses through Stern-Gerlach experiments, which have no further immediate conceptual use in quantum computing. Instead of relying on these, superposition as a property of a unit vector in a Hilbert space, which is a particular vector space where vectors have a constant $L_2$ norm of one, is focused on.

Another feature of the developed educational material is that it does not involve any complex numbers and is restricted to real numbers, hence not requiring any digression into complex numbers. This hastens the learning process by reducing the level of difficulty the students face. It is possible to discuss most of the introductory topics in quantum computing without referring to complex numbers, and participants that became familiar with these can make the jump to a wider range of subjects using complex numbers as an extension of what they have already learned.

The two main introductory concepts of quantum computing for the gate-based model, which is the predominant model of computation in the field, are qubits and quantum gates. Qubits are the smallest units of quantum states, that can be taken as a generalization of bits. Quantum gates or quantum operators are the allowed operations on these qubits by the laws of quantum physics. Focusing on real numbers, quantum logic gates (which are the fundamental blocks of circuits used for quantum algorithms) are introduced as rotation and reflection operations on the 2D plane, which are represented as unitary matrices. So, for anyone with a background in linear algebra, a quantum state is a vector, and quantum gates are unitary matrices acting upon this state, preserving its norm. Similarly, entangled systems (comprised of multiple qubits with quantum correlations that cannot be described by classical correlations) are just vectors in a higher dimension that cannot be represented as tensor products of single-qubit states, and two-qubit gates are $4$x$4$ unitary matrices. Measurement, which is the only quantum operation that does not preserve the norm of the state vector, is described as the projection operation on this vector to two orthogonal planes, where the associated probabilities are given as the appropriate functions depending on the angle between the state vector and these planes. This procedure does not require any introduction of physical phenomena. Concepts of superposition and entanglement can entirely be explained using algebraic concepts, leaving the route taken through physics rather cumbersome for learners who will not need such understanding in their applications of quantum computing. Additionally, this approach does not restrict further elaboration on the correspondence between these mathematical constructs and physical phenomena but leaves it optional.

\subsection{Content}
The educational material utilized in this study is composed of Jupyter notebooks and it runs on the Qiskit framework of IBM. Among the other alternatives such as Cirq, PyQuil, Q\#, and ProjectQ, Qiskit comes forward as IBM Quantum Experience is the first platform to offer publicly available quantum computers and has a large community of users. It uses Python as the primary programming language which also makes it more accessible. Bearing these in mind, Qiskit is chosen as the framework for the content of the educational material developed.

The notebooks in the material can be split into three categories. There are auxiliary notebooks that should be completed by the participants before participating the workshop, reviewing basic linear algebra and basics of Python programming. There are reference notebooks on Python and Qiskit, to guide the participants whenever needed. Finally, there are task-based learning notebooks which are a combination of explanations followed by small sections of coding tasks.

The section reviewing basic linear algebra provides information on vectors, dot product, matrices, and tensor product. Any support library (such as Numpy) is not used for linear algebraic operations as a pedagogical strategy, scaffolding participants to perform matrix-vector multiplications by hand to understand the evolution of states through linear operators. Python review consists of basic structures such as variables, loops, conditionals, and lists. Reference notebooks contain all sufficient information about Qiskit and Python for the participants to complete the entire educational material.

At the core of the material lie around 30 notebooks which cover the following topics:

\begin{itemize}
\item Basics of Classical Systems
\item Basics of a Quantum Program
\item Basics of Quantum Systems
\item Quantum Operators on a Real-Valued Qubit
\item Quantum Correlation
\item Grover's Search Algorithm
\end{itemize}

Each notebook consists of some theoretical background about the subject, programming, and conceptual tasks, accompanied by a notebook in which the solutions for the tasks can be found. Some of the first tasks involve only Python programming, and the Qiskit library is introduced gradually as new quantum computing concepts come forward. Most of the conceptual tasks involve mathematical derivations which should be completed on paper. Overall, the focus is on practical concepts and quantum programming instead of theoretical proofs. Aimed to serve as introductory material for a three-day-long workshop, some traditional topics usually included in quantum computing courses, such as Deutsch and Bernstein-Vazirani algorithms, are purposefully omitted.

The first section of the material introduces the basics of classical systems. The section starts with classical bits and concentrates on probabilistic bits and states. Vector representation for probabilistic states and their evolution through operators by vector-matrix multiplication is given through a game of coin flips. A notation for probabilistic states is also introduced which builds the foundations of the bra-ket or Dirac notation, which is a widely used formalism in quantum mechanics. The tasks in this section do not involve quantum programming but instead include random probabilistic state generation or simulation of a coin-flip.

Before moving into the basics of quantum systems, a notebook describing how to run quantum circuits in Qiskit is provided. This is followed by the section about the basics of quantum systems. The section starts with an interferometer setup which is presented as a quantum coin-flipping experiment, the quantum analog of the coin-flipping previously introduced in the basics of the classical systems section. The following notebooks contain tasks that ask for the implementation of the experimental setups in Qiskit. The section also contains notebooks on quantum states, quantum operators, and superposition and measurement. 

The following section is about quantum operators and their visualization. As the scope of the educational material is restricted to real numbers, the quantum operators can be represented in the 2D plane as reflections and rotations. The tasks in this section include drawing random quantum states on the 2D plane using Python and involve both encoding and visualization of these states.

The next subjects introduced are multiple qubits and entanglement. The participants are already familiar with concepts like the tensor product and the notation for multiple qubits, as they were introduced to in the probabilistic systems section. Superdense coding and quantum teleportation protocols, which are communication protocols that make use of the properties like entanglement, are continually discussed alongside entanglement. Tasks in the section ask for the implementation of both protocols. There also some tasks in which mathematical derivations should be performed such as verification of the superdense coding protocol.

The last section of the material is on Grover's Search algorithm, a quantum database search algorithm that provides quadratic speedup against corresponding classical database search algorithms\cite{grover2001} and almost all previous content builds the foundations for this section. Each iteration of the algorithm consists of two phases called the query and the inversion phases and the consecutive application of those two phases increases the probability of measuring the searched element which is known as the amplitude amplification procedure. The first notebook aims to introduce amplitude amplification idea through an ``Inversion about the mean'' game, where the idea of query and inversion phases are applied on a list of elements and the change of the amplitudes along with Grover's iterations are simulated visually. One qubit representation of Grover's Search is presented next, which helps visualizing the whole algorithm and finally implementation of the algorithm is discussed.

\subsection{Survey Design}\label{sec: met}

To assess the knowledge acquisition in participants of workshops on the educational material described above, one group pre/post-test design has been used for the study. The study consists of pre/post-test results of 22 two/three-day workshops organized in 10 countries between May of 2019 and March of 2020 and was finalized due to the global COVID-19 outbreak. Out of 430 participants who have completed the workshops, 317 participants filled both pre-test and post-tests.

Although the same material was conducted for each workshop, instructors and mentors were different. The workshops were free of charge, organized by volunteering members and they were open to all willing participants. The number of participants was limited to at most 40 for an effective classroom environment, where mentors can assist participants. For the workshops with more than 40 applicants, a selection process was implemented. As a first rule, applicants from disadvantaged minorities and women were prioritized, followed by an assessment of motivation (which was asked in the application form). Participants with no answers or very brief and generic responses like ``I am curious'' for the motivation question were not accepted in such cases. No restrictions on department or education level were imposed for participants, which allowed people from different fields to participate in the workshops.

At the beginning of each workshop participants were provided with ID numbers for them to mark their pre/post-test to allow individual comparison while allowing anonymity. The organizers did not record these participant IDs. An informed consent form was provided with both tests, and it is also verbally noted that participation is voluntary. Ratio of the participants who filled both tests out of the total set of participants is 74\%. The form was generated using Google Forms. Gender, age, education level, and departments (current or graduated) were asked as demographic data, and only the department information was a required section to submit the test. Hence any participant that might not want to provide their demographic data in terms of gender, age, or education level was respected. A summary of the demographics information is given in Table \ref{table: demo}.

\begin{table}[h]
\centering

\caption{Demographics information of the participants}
\label{table: demo}
\centering
\begin{tabular}{@{}lcllcllc@{}}
\toprule
Gender & $n$                    &  & Education & $n$                    &  &     Age  & $n$   \\ \midrule
Female                    & 83                   &  & High School                 & 19                   &  & 14-17 & 11  \\
Male                    & 227                  &  & Bachelors                 & 144                  &  & 18-24 & 171 \\
                     &                      &  & Master's                 & 75                   &  & 25-34 & 87  \\
                     &                      &  & Ph.D.                & 71                   &  & 35-44 & 26  \\
                     & \multicolumn{1}{l}{} &  &                      & \multicolumn{1}{l}{} &  & 45-60 & 11  \\ \bottomrule
\end{tabular}
\end{table}

The purpose of this survey was to measure knowledge acquisition, which according to Bloom's Taxonomy of Educational Objectives is the most basic level \cite{BEFH56}. The revised taxonomy by Anderson \cite{anderson99}, replaces the term knowledge with remember. In \cite{eber07}, this level is described as involving learning facts, knowledge of major ideas and memorizing and it is claimed that it builds a foundation for the remaining levels of cognition, which aligns with the aims of this study. 

The questions were developed by several organizers together to test whether the participants were learning the essential points of the material. There are seven questions, one on the base programming language that Qiskit runs on (which is Python), and six on the concepts of quantum computing such as quantum logic gates, qubits, teleportation, superdense coding, and Grover’s algorithm. The list of questions can be found in Table \ref{table: questions}. The same questions were asked on both the pre/post-test. The demographic data is collected with the pre-test. In the post-test, the attendance level of the participants, satisfaction questionnaire with a Likert type scale, and feedback on the workshop were also asked.

\begin{table}[h]
\centering
\caption{Graded questions on the pre/post-test.}
\label{table: questions}
{\begin{tabular}{p{5.5cm}c}\toprule
 Questions & Answer Type\\ 
 \midrule
What are the two fundamental quantum phenomena that differentiates quantum computing from classical computing? & Checklist   \\
What is the programming language that Qiskit runs on? & Multiple Choice    \\
Match the  quantum  logic  gates  with  their  respective  matrix  representations (leave empty if you don’t know the subject). & Matching \\
Which of the following elements are not necessary for quantum teleportation? Select all that applies. & Checklist \\
Which quantum resource is used for superdense coding? & Multiple Choice \\
What is the common property of probabilistic bits and qubits? & Multiple Choice \\
What is Grover’s algorithm used for? & Multiple Choice \\
 \bottomrule
\end{tabular}}
\end{table}

\subsection{Assessment of Knowledge Acquisition}
For each participant, basic knowledge acquisition is measured by the gain score, which is the difference between the pre/post-test scores.

Normalized gain score is a measure of change to assess the knowledge of the students at the beginning and at the end of a course when the same test is used. According to \cite{Ha98}, normalized gain score ($n_{gain}$) is calculated as

$$
 \frac{M_{post} - M_{pre}}{100- M_{pre}},
$$
where $M_{pre}$ and $M_{post}$ are the average pre/post-test scores respectively. The $n_{gain}$ score is interpreted as high if $n_{gain}$ $>$ 0.7, medium if 0.7 $>$ $n_{gain}$ $>$ 0.3, and low if $n_{gain}$ $<$ 0.3.

An alternative method to measure the magnitude of the difference between two groups is the effect size. A metric proposed for the effect size is Cohen's $d$ index \cite{Co88} which is calculated as follows:

$$
\frac{M_{post}-M_{pre}}{SD_{p}} \mbox{ where } SD_{p}=\sqrt{\frac{SD_{post}^2 + SD_{pre}^2}{2}}
$$
and $SD_{post}$ and $SD_{pre}$ are the standard deviations of the post and pre-test scores respectively.

A $d$ value less than 0.3 is considered as a small; a $d$ value between 0.3 and 0.6 is considered as a moderate, and a $d$ value larger than 0.6 is considered as a large effect size.

\section{Findings}\label{sec: res}

\subsection{Survey Results}

In this section, the main findings of the analyses performed on the pre/post-test data are provided. An alpha level of $.05$ for all statistical tests was used.

Normality of the gain scores ($M=38.77,~SD=24.21$) of the 317 participants is tested with D'Agostino Pearson test as the sample size is large ($>300$) \cite{DBD90} and the test reveals that the gain scores are distributed normal ($p$=0.11).

Paired $t$-test is used in order to compare the pre-test ($M=32.32,~SD=25.03$) and post-test ($M=71.08,~SD = 20.83$) scores. The results indicate a significant difference in the gain scores of the participants. $n_{gain}$ score is calculated as 0.56, suggesting a medium gain. The effect size for the analysis ($d = 1.68$) was found to be large according to Cohen's convention. 

\begin{table}[H]
\centering 
\caption{$t$-test results for the gain score.}
\begin{tabular}{@{}lcccc@{}}
\toprule
\multicolumn{2}{l}{$t$-test} &       \multicolumn{2}{c}{$\% 95$ CI for gain}  & \\ \midrule
$t$           & df           & Lower & Upper              & Cohen's $d$              \\ \midrule 
28.51            & 316          & 36.09 & 41.45              & 1.68                     \\ \bottomrule
\end{tabular}
\end{table}

Considering the 258 participants who provided information about their current or graduated departments, these are gathered into six categories: computer science, engineering,  physics, science, and high school. Engineering category consists of the students with backgrounds in electrical and electronic engineering and other engineering departments. Those who do not study physics, computer science and engineering but a science-related department is categorized as science. The 7 participants who don't have an educational background in a STEM field are categorized as other and are not listed under the results as the sample size is small. Results are summarized in Table \ref{table: dep}.

\begin{table}[]
\centering
\caption{Difference with respect to educational background.}
{\begin{tabular}{@{}llrlrlrlc@{}}
\toprule
               &    & \multicolumn{2}{l}{Pre-test}                          & \multicolumn{2}{l}{Post-test}                         &       \multicolumn{2}{l}{Gain}                             \\ \midrule
               Dep.     & n  & \multicolumn{1}{l}{Mean} & SD                     & \multicolumn{1}{l}{Mean} & SD                     & \multicolumn{1}{l}{Mean } & SD  & $n_{gain}$                \\ \midrule 
CS             & 91  & 30.00                 & 20.24                    & 73.41                & 18.29                    & 43.41                 &22.49   &0.62        \\ 
PHYS           & 62 & 39.16                 & \multicolumn{1}{r}{25.80} & 78.27                & \multicolumn{1}{r}{18.87} & 39.11                      & 26.54 & 0.64 \\ 
ENG   & 47 & 29.26                & 25.28                     & 68.02                & 21.26             & 38.77                      & 26.66           & 0.55  \\
SCI     & 32 & 28.75                   & {18.76} & 68.78                   & 20.24                     & 40.03                         & 18.72 & 0.56 \\
HS & 19  & 12.89             & 12.76                    & 59.11                & \multicolumn{1}{r}{22.79} & 46.21                      & 21.76 & 0.53  \\   
\bottomrule
\end{tabular}}
\label{table: dep}
\end{table}

For each group, Shapiro-Wilk test was conducted which revealed that the gain scores are distributed normal. Consequently, one-way ANOVA test is conducting suggesting that there is no significant difference between the gain scores of the participants with different educational backgrounds, $F(4,246) = .66, p=.62$.

\begin{table}[H]
\centering
\caption{$t$-test results for different educational backgrounds.}
\begin{tabular}{@{}llllll@{}}
\toprule
            & \multicolumn{2}{l}{$t$-test} & \multicolumn{2}{c}{$\% 95$ CI for gain} &             \\
            \midrule
Dep.  & $t$              & df        & Lower                 & Upper                 & Cohen's $d$ \\ \midrule
CS          & 18.41*      & 90        & 38.72                 & 48.09                 & 2.25       \\
ENG         & 9.97*         & 46        & 30.94                 & 46.59                 & 1.66        \\
PHYS        & 11.61*        & 61        & 32.37                 & 45.85                 & 1.72        \\
SCI         & 12.10*        & 31        & 33.28                 & 46.78                 & 2.02        \\
HS & 9.26*          & 18        & 35.72                 & 56.70                  & 2.44        \\ \bottomrule
\end{tabular}
\\
\vspace{0.1in}

Note: $ ^*p < .001$
\label{table: dep2}
\end{table}

In Table \ref{table: dep2}, the results of the $t$-test conducted for each group are presented. It is observed that there is a significant increase in the gain scores for all groups and a large effect size.  

\begin{figure}[H]
\centering
\includegraphics[width=0.5\textwidth]{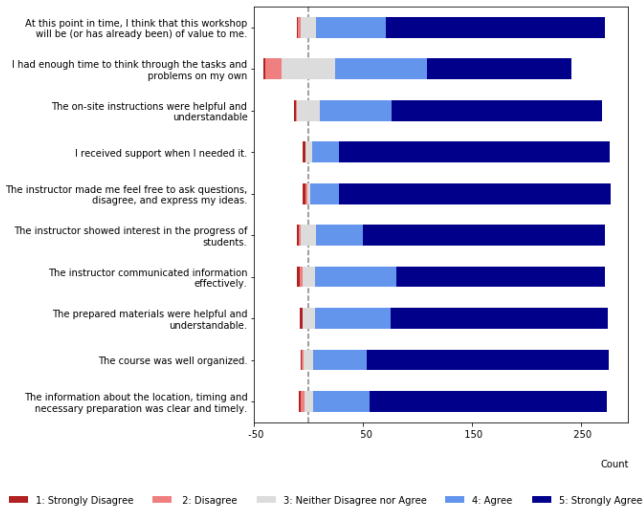}
\caption{Questions and results of the satisfaction test attached to the post-test.}
\label{fig: sats}
\end{figure}

Post-test consisted of 10 statements related to participant satisfaction measured on a 5-point Likert scale ranging from 1 (strongly disagree) to 5 (strongly agree). Responses of the 283 participants for each statement are visualized in Figure \ref{fig: sats} using a stacked bar chart. The participants who agree
with the statement are shown to the right of the zero line whereas those who disagree
are shown to the left. The participants who neither agree nor disagree are
split down the middle.

The averages were above 4.5 for each statement except statement 2 which had an average score of 4.17. The median value was 5 for each question. 

\section{Discussion}
\subsection{Knowledge Retention}

The study aimed to share a methodology in introducing quantum computing and to assess the knowledge retention in basic quantum computing concepts after completing a material prepared by this mindset. Depending on our findings presented in the previous section, several points can be discussed.

The material has been effective in terms of increasing the basic knowledge level of the participants. Both the $n_{gain}$ score and effect size support this claim. However, it is important to bear in mind that the knowledge level mentioned here corresponds to the lowest level in Bloom's Taxonomy as discussed in Section \ref{sec: met}. 

Participants with a physics background have the highest mean pre-test scores, which might be explained by the fact that it is more likely that they were acquainted with quantum computing before compared to other groups. When the effect size is checked, it can be seen that the participants who are high school students and the ones with computer science backgrounds have higher effect sizes. 

No significant difference in the gain scores was found between the participants with different backgrounds. This was expected as the material was aimed to be beneficial for a diverse audience, not for a specific group. A significant increase in the gain scores was observed for all the groups. It can be concluded that the material has been effective in increasing the basic knowledge level of the participants with diverse backgrounds mainly rooted in STEM fields. Even though no background information is known for high school students, one might expect them to come from STEM backgrounds as well due to programming and linear algebra prerequisites.

\subsection{Satisfaction Survey}

Participants answered 10 questions about their satisfaction levels. One of the questions asked whether the workshop has already been of value for them, and a vast majority strongly agrees with this claim. The lowest mean score is found for the second question, which asks whether the participants had enough time to complete the tasks on their own. This is understandable as the material is quite dense for 3 days, and sometimes participants struggle to complete the tasks.

\subsection{Threats to Validity}\label{sec: threats}

This subsection lists the threats to the validity of the survey conducted. Validity can be divided into four categories \cite{WHH03}: internal validity, external validity, conclusion validity, construct validity. Next, each one will be explored in more detail.

In single group pre/post design, some of the most common threats to internal validity are listed as history, maturity, instrumentation, statistical regression, and attrition \cite{PM04}. Workshops were organized either on two or three consecutive days and continued the whole day, not leaving room for any other intervention that might result in history and maturation threat. Instrumentation threat neither poses a challenge, as the performance evaluation metric relies on objective measures. Attrition, which is also known as mortality, takes place when some of the participants drop out of the experiment. In this setting, out of 430 participants who completed the workshops, 421 participants answered the pre-test, 317 participants answered both tests, and the test results of the participants who only answered one of the tests are discarded. When the mean score of the pre-tests with and without discarding the ones who did not answer the post-test are compared, the mean scores are 32.32 and 31.11 respectively, not yielding any implication about the pre-knowledge of the participants who did not answer post-test.

Even though environmental factors may pose a threat to external validity, the data analyzed was collected from 22 workshops organized in 10 countries by different educators and mentors. The repetitive nature of the data is evidence of the generalizability of these results to a certain extend. A possible threat might be volunteer bias, which stems from the fact that filling the pre/post-test surveys was not compulsory. Previous studies evaluating volunteer bias have reported that volunteers differ from the rest of the sample in various means \cite{RR76}, and one might argue that the participants who did not choose to answer are the ones with lower success rates.

The design of the test questions together with the gain scores as the performance metric can be a threat to construct validity, as the experiment is designed as a one-group pre/post-test design. The normalized gain scores are considered together with the gain scores to overcome the ``You can only go up from here'' phenomenon. Cohen's $d$ metric is also taken into consideration to measure the effect size. Nevertheless, further attention can be paid to assess the reliability of the test questions by using various statistical tests \cite{WW03}. 

Conclusion validity is concerned with the ability to draw correct conclusions based on the outcomes of the experiment. For hypothesis testing to decide whether there was a significant difference between pre/post-test scores, paired $t$-test is used after testing the normality of the data. To see whether there is a significant difference between the gain scores of the participants from different groups, the normality test was conducted followed by two-sample tests, Kruskal Wallis and one-way ANOVA, depending on the structure of the data.

\section{Conclusion} \label{sec: con}

As the field of quantum computing emerges rapidly, and the need for a quantum workforce for industry grows urgent, expanding beyond the roots in physics is becoming a necessity for educational purposes. The purpose of the current study was to inform academics and organizations interested in introducing quantum computing to a diverse group of participants on a possible educational approach for quantum computation, originating from computer science-oriented thinking\cite{YM08, AB09, aaronson2013}.

It is argued that this approach is more appealing for an audience who does not want to learn quantum physics or will not need such understanding, and facilitate increasing quantum awareness \cite{Laur} among a diverse group.  Being introduced to quantum programming is another educational attainment of the approach, contributing towards the endeavor to prepare for the future quantum workforce \cite{FZL20,Pe20,qusoft_manf,gerke2020}.

The viability of this approach is demonstrated through pre/post-test analysis, and the reported satisfaction levels of the participants for 22 workshops organized in 10 countries. Outcomes revealed a significant increase in the scores for participants from diverse backgrounds. Since the topic is relatively recent and there are only a few empirical studies\cite{MS20,HIPST20,PSHIT19,larose2019}, which does not allow for comparison with the type of detailed analysis provided here, these results can be accepted as preliminary findings. In spite of its limitations, this work offers valuable insights for the field of quantum computing education.

Following the large-scale public funding schemes discussed previously \cite{q_us1,q_us2,q_eu}, there are new organizational and collaborative efforts in quantum education, such as the National Q-12 Education Partnership\footnote{https://q12education.org/. Accessed: May 4, 2021} in the US, and the QTEdu\footnote{https://qt.eu/about-quantum-flagship/projects/education-coordination-support-actions/. Accessed: May 4, 2021} in the EU. Through the development of these and similar initiatives, research on comparative advantages of traditional (physics-based) approach and computer science-oriented approach to knowledge acquisition in quantum computing under different conditions (class environment, workshops, MOOCs) is most likely to increase. The methods and findings provided in this work can be utilized by these studies.

Similarly, the content provided for the educational material described in this study can be utilized by academics and organizations for their curriculum development and content generation processes. The educational material (called Bronze) has been used in more than 50 workshops in 20 countries to date\footnote{https://qworld.lu.lv/index.php/workshop-bronze/. Accessed: May 4, 2021}, and it is an open-source material. 

Further investigation into this research can be pursued in several ways. First, this study can be re-iterated with multiple groups, utilizing two similar educational materials developed adopting different approaches (physics-based and computer science-oriented). Such a study requires access to a wide pool of participants, but collaborative structures such as National Q-12 Education Partnership make this an achievable goal. Secondly, experimental studies can be performed by developing and implementing materials adopting this approach for different mediums (such as MOOCs). Finally, assessment methods can be switched and instead of pre/post-test surveys, alternative methods might be adopted (such as qualitative means like one-on-one interviews, focus group interviews, journal keeping). Mixed-method approaches that combine qualitative and quantitative methods would also reveal insights into the learning processes of learners introduced to the topic. A side question raised by this study in need of further investigation is how much quantum mechanics knowledge one needs to enter into the field of quantum computing. 

Quantum computing is a relatively new field; quantum programming is even newer. Therefore, it is safe to assume that there is still much to learn about how to best adjust the educational materials for different audiences, mediums, and teaching methods. Exploring such aspects can yield valuable outcomes that can be utilized in the path forward to the quantum era.

\section*{Acknowledgment}

The analysis parts of this work has been completed under the QIntern program organized by QWorld, therefore the authors would like to thank to the organizers of the program. 

Additionally, we would like to extend our gratitude to all the organizers of workshops that help administer the pre/post-tests. We thank Dr. Ula\c{s} \.{I}lic and Franziska Gerke for their feedback and suggestions. Finally, we would like to thank Dilan K\"{o}se, Dr. Abuzer Yakary{\i}lmaz, and Sercan Erer for their invaluable contributions to the study.

\bibliographystyle{IEEEtran}
\bibliography{references.bib}

\end{document}